\documentclass[aps,prl,twocolumn,showpacs,groupedaddress]{revtex4}  
\usepackage{graphicx}  
\usepackage{dcolumn}   
\usepackage{bm}        
\usepackage{amssymb}   

\begin{document}
\def\et{$E_T$}                          
\def\et{$p_T$}                          
\def\D0{D\O}                            
\def\ppbar{$p\overline{p} $}            
\def\ttbar{$t\overline{t} $}            
\def\pbarp{$\overline{p}p $}            
\def\pb{$pb^{-1}$}
\def\eebar{$e^{+}e^{-}$}
\def\etal{{\sl et al.\ }}
\def\zjet{$Z+${\rm jets}}
\newcommand{\etadet}{\ensuremath{\eta_{\mrm det}}}
\newcommand{\met}       {\mbox{$\not\!\!E_T$}}
\newcommand{\mrm}{\mathrm}
\newcommand{\etopt}{\ensuremath{E_T/p_T}}
\newcommand{\wenu}{\ensuremath{W{\rightarrow}e^\pm\nu}}
\newcommand{\zee}{\ensurematch{Z \rightarrow e^{+}e^{-}}}
\newcommand{\pythia}    {\sc{pythia}}
\newcommand{\alpgen}    {\sc{alpgen}}
\newcommand{\geant}    {\sc{geant}}
\newcommand{\madgraph}    {\sc{madgraph}}
\newcommand{\mcfm}    {\sc{mcfm}}

\title{Measurement of the ratios of the $Z/\gamma^* + \geq n$ jet production cross
sections to the total inclusive $Z/\gamma^*$ cross section in
\ppbar\ collisions at $\sqrt{s}$ = 1.96 TeV}

%
\author{                                                                      
V.M.~Abazov,$^{36}$                                                           
B.~Abbott,$^{76}$                                                             
M.~Abolins,$^{66}$                                                            
B.S.~Acharya,$^{29}$                                                          
M.~Adams,$^{52}$                                                              
T.~Adams,$^{50}$                                                              
M.~Agelou,$^{18}$                                                             
S.H.~Ahn,$^{31}$                                                              
M.~Ahsan,$^{60}$                                                              
G.D.~Alexeev,$^{36}$
% ...
\\                                                                            
\vskip 0.30cm                                                                 
\centerline{(D\O\ Collaboration)}                                             
\vskip 0.30cm                                                                 
}                                                                             
\affiliation{                                                                 
\centerline{$^{1}$Universidad de Buenos Aires, Buenos Aires, Argentina}       
\centerline{$^{2}$LAFEX, Centro Brasileiro de Pesquisas F{\'\i}sicas,         
                  Rio de Janeiro, Brazil}                                     
\centerline{$^{3}$Universidade do Estado do Rio de Janeiro,                   
                  Rio de Janeiro, Brazil}                                     
\centerline{$^{4}$Instituto de F\'{\i}sica Te\'orica, Universidade            
                  Estadual Paulista, S\~ao Paulo, Brazil}                     
\centerline{$^{5}$University of Alberta, Edmonton, Alberta, Canada,           
                  Simon Fraser University, Burnaby, British Columbia, Canada,}
\centerline{York University, Toronto, Ontario, Canada, and                    
                  McGill University, Montreal, Quebec, Canada}                
\centerline{$^{6}$Institute of High Energy Physics, Beijing,                  
                  People's Republic of China}                                 
\centerline{$^{7}$University of Science and Technology of China, Hefei,       
                  People's Republic of China}                                 
\centerline{$^{8}$Universidad de los Andes, Bogot\'{a}, Colombia}             
\centerline{$^{9}$Center for Particle Physics, Charles University,            
                  Prague, Czech Republic}                                     
\centerline{$^{10}$Czech Technical University, Prague, Czech Republic}        
\centerline{$^{11}$Center for Particle Physics, Institute of Physics,         
                   Academy of Sciences of the Czech Republic,                 
                   Prague, Czech Republic}                                    
\centerline{$^{12}$Universidad San Francisco de Quito, Quito, Ecuador}        
\centerline{$^{13}$Laboratoire de Physique Corpusculaire, IN2P3-CNRS,         
                   Universit\'e Blaise Pascal, Clermont-Ferrand, France}      
\centerline{$^{14}$Laboratoire de Physique Subatomique et de Cosmologie,      
                   IN2P3-CNRS, Universite de Grenoble 1, Grenoble, France}    
\centerline{$^{15}$CPPM, IN2P3-CNRS, Universit\'e de la M\'editerran\'ee,     
                   Marseille, France}                                         
\centerline{$^{16}$IN2P3-CNRS, Laboratoire de l'Acc\'el\'erateur              
                   Lin\'eaire, Orsay, France}                                 
\centerline{$^{17}$LPNHE, IN2P3-CNRS, Universit\'es Paris VI and VII,         
                   Paris, France}                                             
\centerline{$^{18}$DAPNIA/Service de Physique des Particules, CEA, Saclay,    
                   France}                                                    
\centerline{$^{19}$IPHC, IN2P3-CNRS, Universit\'e Louis Pasteur, Strasbourg,  
                    France, and Universit\'e de Haute Alsace,                 
                    Mulhouse, France}                                         
\centerline{$^{20}$Institut de Physique Nucl\'eaire de Lyon, IN2P3-CNRS,      
                   Universit\'e Claude Bernard, Villeurbanne, France}         
\centerline{$^{21}$III. Physikalisches Institut A, RWTH Aachen,               
                   Aachen, Germany}                                           
\centerline{$^{22}$Physikalisches Institut, Universit{\"a}t Bonn,             
                   Bonn, Germany}                                             
\centerline{$^{23}$Physikalisches Institut, Universit{\"a}t Freiburg,         
                   Freiburg, Germany}                                         
\centerline{$^{24}$Institut f{\"u}r Physik, Universit{\"a}t Mainz,            
                   Mainz, Germany}                                            
\centerline{$^{25}$Ludwig-Maximilians-Universit{\"a}t M{\"u}nchen,            
                   M{\"u}nchen, Germany}                                      
\centerline{$^{26}$Fachbereich Physik, University of Wuppertal,               
                   Wuppertal, Germany}                                        
\centerline{$^{27}$Panjab University, Chandigarh, India}                      
\centerline{$^{28}$Delhi University, Delhi, India}                            
\centerline{$^{29}$Tata Institute of Fundamental Research, Mumbai, India}     
\centerline{$^{30}$University College Dublin, Dublin, Ireland}                
\centerline{$^{31}$Korea Detector Laboratory, Korea University,               
                   Seoul, Korea}                                              
\centerline{$^{32}$SungKyunKwan University, Suwon, Korea}                     
\centerline{$^{33}$CINVESTAV, Mexico City, Mexico}                            
\centerline{$^{34}$FOM-Institute NIKHEF and University of                     
                   Amsterdam/NIKHEF, Amsterdam, The Netherlands}              
\centerline{$^{35}$Radboud University Nijmegen/NIKHEF, Nijmegen, The          
                  Netherlands}                                                
\centerline{$^{36}$Joint Institute for Nuclear Research, Dubna, Russia}       
\centerline{$^{37}$Institute for Theoretical and Experimental Physics,        
                   Moscow, Russia}                                            
\centerline{$^{38}$Moscow State University, Moscow, Russia}                   
\centerline{$^{39}$Institute for High Energy Physics, Protvino, Russia}       
\centerline{$^{40}$Petersburg Nuclear Physics Institute,                      
                   St. Petersburg, Russia}                                    
\centerline{$^{41}$Lund University, Lund, Sweden, Royal Institute of          
                   Technology and Stockholm University, Stockholm,            
                   Sweden, and}                                               
\centerline{Uppsala University, Uppsala, Sweden}                              
\centerline{$^{42}$Physik Institut der Universit{\"a}t Z{\"u}rich,            
                   Z{\"u}rich, Switzerland}                                   
\centerline{$^{43}$Lancaster University, Lancaster, United Kingdom}           
\centerline{$^{44}$Imperial College, London, United Kingdom}                  
\centerline{$^{45}$University of Manchester, Manchester, United Kingdom}      
\centerline{$^{46}$University of Arizona, Tucson, Arizona 85721, USA}         
\centerline{$^{47}$Lawrence Berkeley National Laboratory and University of    
                   California, Berkeley, California 94720, USA}               
\centerline{$^{48}$California State University, Fresno, California 93740, USA}
\centerline{$^{49}$University of California, Riverside, California 92521, USA}
\centerline{$^{50}$Florida State University, Tallahassee, Florida 32306, USA} 
\centerline{$^{51}$Fermi National Accelerator Laboratory,                     
            Batavia, Illinois 60510, USA}                                     
\centerline{$^{52}$University of Illinois at Chicago,                         
            Chicago, Illinois 60607, USA}                                     
\centerline{$^{53}$Northern Illinois University, DeKalb, Illinois 60115, USA} 
\centerline{$^{54}$Northwestern University, Evanston, Illinois 60208, USA}    
\centerline{$^{55}$Indiana University, Bloomington, Indiana 47405, USA}       
\centerline{$^{56}$University of Notre Dame, Notre Dame, Indiana 46556, USA}  
\centerline{$^{57}$Purdue University Calumet, Hammond, Indiana 46323, USA}    
\centerline{$^{58}$Iowa State University, Ames, Iowa 50011, USA}              
\centerline{$^{59}$University of Kansas, Lawrence, Kansas 66045, USA}         
\centerline{$^{60}$Kansas State University, Manhattan, Kansas 66506, USA}     
\centerline{$^{61}$Louisiana Tech University, Ruston, Louisiana 71272, USA}   
\centerline{$^{62}$University of Maryland, College Park, Maryland 20742, USA} 
\centerline{$^{63}$Boston University, Boston, Massachusetts 02215, USA}       
\centerline{$^{64}$Northeastern University, Boston, Massachusetts 02115, USA} 
\centerline{$^{65}$University of Michigan, Ann Arbor, Michigan 48109, USA}    
\centerline{$^{66}$Michigan State University,                                 
            East Lansing, Michigan 48824, USA}                                
\centerline{$^{67}$University of Mississippi,                                 
            University, Mississippi 38677, USA}                               
\centerline{$^{68}$University of Nebraska, Lincoln, Nebraska 68588, USA}      
\centerline{$^{69}$Princeton University, Princeton, New Jersey 08544, USA}    
\centerline{$^{70}$State University of New York, Buffalo, New York 14260, USA}
\centerline{$^{71}$Columbia University, New York, New York 10027, USA}        
\centerline{$^{72}$University of Rochester, Rochester, New York 14627, USA}   
\centerline{$^{73}$State University of New York,                              
            Stony Brook, New York 11794, USA}                                 
\centerline{$^{74}$Brookhaven National Laboratory, Upton, New York 11973, USA}
\centerline{$^{75}$Langston University, Langston, Oklahoma 73050, USA}        
\centerline{$^{76}$University of Oklahoma, Norman, Oklahoma 73019, USA}       
\centerline{$^{77}$Oklahoma State University, Stillwater, Oklahoma 74078, USA}
\centerline{$^{78}$Brown University, Providence, Rhode Island 02912, USA}     
\centerline{$^{79}$University of Texas, Arlington, Texas 76019, USA}          
\centerline{$^{80}$Southern Methodist University, Dallas, Texas 75275, USA}   
\centerline{$^{81}$Rice University, Houston, Texas 77005, USA}                
\centerline{$^{82}$University of Virginia, Charlottesville,                   
            Virginia 22901, USA}                                              
\centerline{$^{83}$University of Washington, Seattle, Washington 98195, USA}  
}                                                                             

\date{March 30, 2007}

\begin{abstract}
We present a study of events with $Z$ bosons and associated 
jets produced at
the Fermilab Tevatron Collider in $p\bar{p}$ collisions at a
center of mass energy of 1.96 TeV. The data sample consists of
nearly 14,000 $Z/\gamma^* \rightarrow e^{+}e^{-}$ candidates
corresponding to an integrated luminosity of 0.4~fb$^{-1}$
collected with the D\O\ detector. Ratios of the $Z/\gamma^* +
\geq n$ jet cross sections to the total inclusive $Z/\gamma^*$
cross section have been measured for $n = 1$ to $4$ jets, and
found to be in good agreement with a
next-to-leading order QCD calculation and with a tree-level QCD
prediction with parton shower simulation and hadronization.
\end{abstract}

\maketitle

Leptonic decays of electroweak gauge bosons, $W^{\pm}$ and
$Z$, produced in association with jets are prominent signatures at
present and future hadron colliders.  Measurements of $W$ (or $Z$) $+\ge
n$~jet cross sections are important for understanding perturbative
quantum chromodynamics (QCD) calculations and for developing Monte Carlo (MC)
simulation programs capable of handling partons in the final
state at leading order (LO), or, in some cases, next-to-leading
order (NLO).  Furthermore, the production of $W$ or $Z$
bosons with associated jets represents a significant background to Higgs boson
searches, as well as to other standard model processes of interest,
such as top quark production, and many searches for new phenomena at the
Fermilab Tevatron Collider and at the CERN Large Hadron Collider.

Measurements of $Z~+ \ge n$~jet cross sections with lower
integrated luminosity and at lower center of mass energy were
performed previously by the CDF collaboration~\cite{cdfrun1}. In
this study, we present the first measurement of the fully corrected ratios of the
$Z/\gamma^* + \geq n$ jet production cross sections to the total
inclusive $Z/\gamma^*$ cross section for jet multiplicities
$n=1-4$ in $p\overline{p}$ collisions at $\sqrt{s}=1.96$~TeV.
Cross section measurements based on inclusive jet multiplicities
provide theoretically sound observables, and can be compared
to a variety of predictions.
Our results are based on a data sample corresponding to an
integrated luminosity of 0.4~fb$^{-1}$ accumulated with the D\O\
detector.

The elements of the \D0\ detector~\cite{run2det} of primary
importance to this analysis are the uranium/liquid-argon sampling
calorimeter and the tracking system. The \D0\ calorimeter has a
granularity of $\Delta\eta\times\Delta\phi = 0.1\times0.1$, forming
projective towers, where $\eta$ is the pseudorapidity ($\eta=-
\ln[\tan(\theta/2)]$, $\theta$ is the polar angle relative to
the proton beam), and $\phi$ is the azimuthal angle. The
calorimeter has a central section covering pseudorapidities up to
$\approx 1.1$, and two end calorimeters that extend the coverage
to $|\eta|\approx 4.2$. The tracking system consists of a silicon
micro-strip tracker and a central fiber tracker, both located
within a 2~T superconducting solenoidal magnet, with designs
optimized for tracking and vertexing at pseudorapidities of
$|\eta|<3$ and $|\eta|<2.5$, respectively.

The data sample for this analysis~\cite{buehler} was collected
between April 2002 and June 2004. Events from $Z/\gamma^*
\rightarrow e^+e^-$ decays were selected with a combination of
single-electron triggers, based on energy deposited in calorimeter
towers ($\Delta\eta\times\Delta\phi = 0.2\times0.2$).
Final event selection was based on detector performance,
event properties, and electron and jet identification criteria.

Events were required to have a reconstructed primary vertex with a
position along the beam direction within 60~cm of the detector center.
Electrons were reconstructed from electromagnetic (EM) clusters in
the calorimeter using a simple cone algorithm. 
The two electron candidates in the event with the highest transverse momentum
components relative to the beam direction ($p_T$), and both with
$p_{T}>25$~GeV, were used to reconstruct the
$Z$ boson candidate. The two electrons were required to be in the
central region of the calorimeter $|\eta_{\mathrm{det}}| < 1.1$
(pseudorapidity $\eta_{\mathrm{det}}$ is calculated relative to the
center of the detector), and at least one required
to fire the trigger(s) for the event. The electron pair also had
to have an invariant mass consistent with the $Z$ boson mass of 75
GeV $< M_{ee} <$ 105 GeV.

To reduce background (mainly from jets misidentified as
electrons), the EM clusters were required to pass three quality
criteria based on the shower profile: (i) the electron had to
deposit at least 90\% of its energy in the 21-radiation-length
EM calorimeter (ii) the lateral and longitudinal
shape of the energy cluster had to be consistent with those of an
electron, and (iii) the electron had to be isolated from other
energy deposits in the calorimeter, with an isolation fraction
$f_{\mathrm{iso}} < 0.15$. (The isolation fraction is defined as
$f_{\mathrm{iso}}=[E(0.4)-E_{\mathrm{EM}}(0.2)]/E_{\mathrm{EM}}(0.2)$,
where $E(R_{\mathrm{cone}})$ and
$E_{\mathrm{EM}}(R_{\mathrm{cone}})$ are respectively the total and EM energies
within a cone of radius $R_{\mathrm{cone}}=\sqrt{(\Delta\eta)^2 +
(\Delta\phi)^2}$ centered around the direction of the electron.)  Additionally, at
least one of the electrons was required to have a spatially
matched track associated with the reconstructed calorimeter
cluster, and the track momentum had to be consistent with the
energy of the EM cluster.  A total of 13,893 events passed the
selection criteria.

Jets were reconstructed using the ``Run II cone
algorithm"~\cite{jet_cone_alg} that combines particles within a
cone of radius $R_{\mathrm{cone}}=0.5$.  Spurious jets from
isolated noisy calorimeter cells were eliminated through selections
on patterns of jet energy deposition. Jets were required to be consistent
with energy depositions measured at the trigger stage. 
This requirement was introduced to address precision readout noise
problems: The jet energy at the Level 1 trigger tower level was
compared to the jet energy derived from the jet cone algorithm,
which was based on calorimeter cell precision readout.
The transverse momentum of each jet was corrected for multiple \ppbar\
interactions, calorimeter noise, out--of--cone showering effects,
and energy response of the calorimeter as determined from the
missing transverse energy balance of photon--jet events~\cite{jes}. 
Jets were required to have $p_T>20$ GeV and $|\eta| < 2.5$, and were
eliminated if they overlapped with electrons from
$Z$ boson decay within $\Delta R = \sqrt{ (\Delta\eta)^{2} +
(\Delta\phi)^{2} }=0.4$. Small losses of jets resulting from this separation
criterion for electrons from $Z$ boson decays were estimated as a function of
the number of associated jets using a {\pythia}~\cite{pythia}
MC sample.

The jet energy resolutions were derived from a measurement in photon+jet
data for low jet energies and dijet data for higher jet energy values.
Fits to the transverse energy asymetry $[p_{T}(1)-p_{T}(2)]/[p_{T}(1)+p_{T}(2)]$
between the transverse momenta of the back-to-back jets and/or photon
($p_{T}(1)$ and $p_{T}(2)$) were then used to obtain the jet energy resolution
as a function of jet rapidity and transverse energy. The largest
contribution to the jet energy resolution uncertainty was due to
limited statistics in the samples used.

The electron efficiencies for trigger, track matching,
reconstruction, and identification were determined from data,
based on a ``tag-and-probe" method.  $Z$ candidates were selected
with one electron (the tag) satisfying a tighter track-matching
requirement to further reduce background contamination, and
another electron (the probe) with all other criteria applied, except the
one under study. The fraction of events with probe electrons
passing the requirement under study determined the efficiency of a
given criterion. The overall trigger efficiency for $Z$ candidates that
survived the analysis selections was found to be greater than
$99\%$. The electron reconstruction and identification
efficiencies were measured as a function of azimuthal angle and
$p_T$, and the average efficiency was found to be about $89\%$.
The combined spatial and energy track-matching efficiency was
measured to be about $77\%$. The electron reconstruction,
selection, trigger, and track-matching efficiencies were examined
as a function of jet multiplicity.  No significant variations of
the efficiencies were observed, except for the track-matching
efficiency, for which the multiplicity dependence was taken into
account in correcting the data.

The kinematic and geometric acceptance for electrons from
$Z/\gamma^*$ decays in the mass region of 75 GeV $< M_{ee} <$ 105
GeV, for a primary vertex within 60~cm of the detector
center, was determined as a function of jet multiplicity. 
An inclusive {\pythia} sample was used to calculate the acceptance for
the inclusive $Z/\gamma^*$ sample. The {\pythia}
events were weighted so that the $p_{T}$ distribution of the $Z$
boson in the MC agreed with data.  The jet multiplicity
dependence of the acceptance was calculated using a $Z/\gamma^* + n$
parton leading-order generator \cite{alpgen}, with the
evolution of partons into hadrons carried out in {\pythia}. All
the samples were processed through full \D0\ detector
simulation using {\geant}~\cite{geant} and the \D0\
reconstruction software.  The overall dielectron acceptance for
the $Z/\gamma^* +\geq 4$ jet sample was found to be about $30\%$
higher than the acceptance for the $Z/\gamma^*$ inclusive sample,
because events with jets tend to recoil from $Z$ bosons of larger $p_{T}$,
and thereby produce decay products that are more likely to
fall within the geometric acceptance. 

The reconstruction and identification efficiency of jets was
determined from a MC sample with full detector simulation, and
processed through the same programs as the data. A scaling
factor was applied to the MC jets to adjust their reconstruction
and identification efficiency to that of jets in data, using the
``$Z~p_{T}$-balance'' method~\cite{heinmiller}. In events with $Z$
candidates, a search was performed for a recoiling jet opposite 
in azimuth to the $Z$ boson. The probability of finding a
recoiling jet as a function of the $p_{T}$ of the $Z$ was measured in data
and MC. The ratio of these probabilities defined the scaling
factor that was applied to the MC jets. After applying the scaling
factor, the jet reconstruction and identification efficiency was
determined by matching particle-level jets (i.e., jets found from
final-state generator-level particles, after parton hadronization) to calorimeter
jets. The efficiency was parameterized as a function of the $p_{T}$
of the particle-level jet, where the $p_{T}$ values were smeared
with jet energy resolutions observed in data, as measured in three $\eta$
regions of the calorimeter. As a cross check, the scaling factor
determined from the ``$Z~p_{T}$-balance'' method was compared to the
scaling factor obtained for photon+jet events, and found to be consistent
with one another.

The primary background to the $Z/\gamma^*$ dielectron
signal is from multijet production, in which the
jets have a large electromagnetic component or they are
mismeasured in some way that causes them to pass the electron
selection criteria. We refer to this instrumental background
as ``QCD''.
For the $Z/\gamma^* +\geq 0-2$ jet samples, a
convoluted Gaussian/Breit-Wigner function was used to fit the
$Z$ lineshape, and an exponential form was used to account for
both the QCD background and the Drell-Yan ($\gamma^*$) component of the signal.
For the lower statistics $Z/\gamma^* +\geq 3$ jet sample,
the contributions from QCD and Drell-Yan components were estimated
from the side bands of the $Z$ in the dielectron invariant mass spectrum. In each case, a
{\pythia} sample was used to disentangle the QCD component from
the Drell-Yan contribution. The background contribution for the
$Z/\gamma^* +\geq 4$ jet multiplicity sample was estimated by
extrapolating to $n=4$ an exponential fit to the QCD background in the
$0-3$ jet multiplicity bins. The background contribution from 
QCD processes was found to be
$3-5\%$, depending on jet multiplicity.
There are also contributions to
$Z/\gamma^*$ candidates that are not from misidentified
electrons, but correspond to other standard model processes (e.g.,
\ttbar\ production, $Z\rightarrow \tau^{+}\tau^{-}$, $W\rightarrow
e\nu$). These small ($<1\%$) irreducible background contributions were also taken into
account in the analysis.

\begin{table*}
\caption{\label{tab:tab_ratios} Cross-section ratios ($R_{n}$) with
statistical and systematic uncertainties (all $\times 10^{-3}$) for different inclusive
jet multiplicities.}
\begin{ruledtabular}
\begin{tabular}{lcccc}
Multiplicity ($Z/\gamma^* +\geq n$ jets)   & $n\geq 1$   & $n\geq 2$             & $n\geq 3$             & $n\geq 4$\\
\hline

\vspace{-.3cm}
\\

$R_{n}$    & 120.1         & 18.6          & 2.8           & 0.90\\
Total Statistical Uncertainty    & $\pm 3.3$           & $\pm 1.4$           & $\pm 0.56$          & $\pm 0.44$\\
Total Systematic Uncertainty     & $-17.1, +15.6$  & $-5.0, +6.2$  & $-1.06, +1.43$    & $-0.40, +0.48$\\
{} Jet Energy Calibration        & $\pm 11.7$     & $\pm 3.3$      & $\pm 0.74$     & $\pm 0.23$ \\
{} Jet Reconstruction/Identification   & $-7.0, +2.2$  & $-2.9, +4.3$  & $-0.64, +0.82$    & $-0.30, +0.40$ \\
{} Unsmearing Procedure        & $-3.6, +2.2$  & $-1.6, +2.4$  & $-0.24, +0.85$    & $-0.08, +0.09$ \\
{} Jet Energy Resolution       & $-2.7, +3.4$  & $-0.04, +0.13$    & $-0.17, +0.15$  & $-0.03, +0.04$ \\
{} Acceptance                    & $\pm 1.8$      & $\pm 0.7$      & $\pm 0.10$     & $\pm 0.003$ \\
{} Efficiencies (Trigger, EM, Track)         & $\pm 8.5$      & $\pm 1.3$      & $\pm 0.20$     & $\pm 0.07$ \\
{} Electron-Jet-Overlap                  & $\pm 3.2$      & $\pm 0.7$      & $\pm 0.14$     & $\pm 0.05$ \\
\end{tabular}
\end{ruledtabular}
\end{table*}

The cross sections as a function of jet multiplicity were
corrected for jet reconstruction and identification efficiencies,
and for event migration due to the finite jet energy resolution of
the detector.  The correction factors were determined using two
independent MC samples, both tuned to match the
measured inclusive jet multiplicity and jet $p_T$ distributions in
data. The first sample was based on {\pythia} simulations. The
second sample (ME-PS) was based on {\madgraph}~\cite{madgraph} 
$Z/\gamma^* + n$ LO Matrix Element (ME)
predictions, using {\pythia} for parton showering (PS) and
hadronization, and a modified CKKW~\cite{ckkw} method to map
the $Z/\gamma^* + n$ parton event into a parton shower
history~\cite{mrenna}.  The ME-PS predictions relied on
{\madgraph} tree-level processes of up to three partons.
Both these samples contained only particle-level jets (i.e., no
detector simulation). The $p_{T}$ of the jets was smeared with the
jet energy resolutions found in data. Subsequently, some jets were 
removed randomly from the sample, to simulate the measured jet
reconstruction/identification efficiencies. The ratio of the
two inclusive jet multiplicity distributions (the generated
distribution and the one with the jet reconstruction/identification
efficiency and energy resolution applied) determined the unsmearing 
correction factors for a given MC sample.  
The weighted averages of the correction factors
corresponding to the two sets of MC procedures were applied
as a function of jet multiplicity to correct the jet multiplicity
spectrum in data.   The differences between the correction factors for
the two calculations contribute to the systematic uncertainty of the
procedure.  Another source of systematic uncertainty was
determined from a closure test, and was estimated by applying the full
unsmearing procedure to a MC control sample. The unsmearing
correction factors range from $1.11$ to $2.9$ for
$\geq 1$ and $\geq 4$ jets, respectively.

The fully corrected ratios, $R_n$, of the $Z/\gamma^* +\geq n$ jet
production cross sections to the inclusive $Z/\gamma^*$ cross
section
\begin{equation}
    R_{n} \equiv
    { {\sigma(Z/\gamma^* +\geq n~{\rm jets})} \over
{\sigma(Z/\gamma^*) } }
\end{equation}

%

\noindent for the mass region 75 GeV $< M_{ee} <$ 105 GeV are
summarized in Table \ref{tab:tab_ratios}.
%
%
Systematic uncertainties include contributions from jet energy 
calibration corrections, jet reconstruction and identification efficiency,
the unsmearing procedure,
jet energy resolution,
and variations in the acceptance for different event generators.
They also take into account
uncertainties in the variation of efficiencies
for the trigger, electron reconstruction, identification, and track
matching as a function of jet multiplicity,
as well as uncertainties due to the electron-jet overlap correction.
All these
uncertainties are assumed to be uncorrelated, and are added in
quadrature to estimate the total systematic uncertainty.  The
statistical uncertainties include contributions from the number of
candidate events, background estimation, acceptance, efficiencies,
and the unsmearing correction.

\begin{figure}[!]
\includegraphics[scale=0.46]{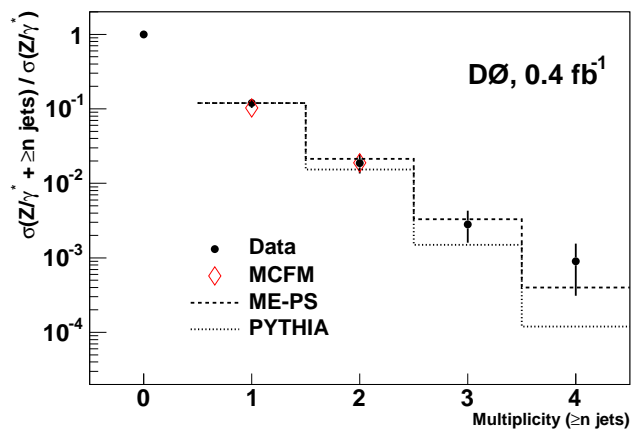}
\caption{\label{fig:fig_ratios} Ratios of the $Z/\gamma^* +\geq n$
jet cross sections to the total inclusive $Z/\gamma^*$ cross
section versus jet multiplicity. The uncertainties on the data
(dark circles) include the combined statistical and
systematic uncertainties added in quadrature. The dashed line
represents predictions of LO Matrix Element (ME) calculations
using {\pythia} for parton showering (PS) and hadronization,
normalized to the measured $Z/\gamma^* +\geq 1$ jet cross-section
ratio.  The dotted line represents the predictions of {\pythia}
normalized to the measured $Z/\gamma^* +\geq 1$ jet cross-section
ratio.  The two open diamonds represent predictions from {\mcfm}.}
\end{figure}

Figure \ref{fig:fig_ratios} shows the fully corrected measured
cross-section ratios for $Z/\gamma^* +\geq n$ jets as a function
of jet multiplicity, compared to three QCD predictions.
{\mcfm}~\cite{mcfm} is a NLO calculation for up to $Z/\gamma^* + 2$
parton processes.  CTEQ6M~\cite{cteq6} parton
distribution functions (PDF) were used in {\mcfm}, and the
factorization and renormalization scales $\mu_{F}$, $\mu_{R}$ were both set to
the $Z$ boson mass, $M_{Z}$. Varying the PDF set and the 
renormalization/factorization scales to $M_{Z}^{2}+p_{T,Z}^{2}$
had a minimal effect on the {\mcfm} cross-section ratios.
The ME-PS predictions are
normalized to the measured $Z/\gamma^* +\geq 1$ jet cross-section
ratio, and use the CTEQ6L PDF, with the
factorization scale set to $\mu_{F}=M_{Z}$, and the
renormalization scale set to $\mu_{R}=p_{Tjet}$ for jets
from initial state radiation and $\mu_{R}=k_{Tjet}$ for jets
from final state radiation ($k_{Tjet}$ is the transverse momentum
of a radiated jet relative to its parent parton momentum). 
The {\pythia} predictions are also normalized to
the measured $Z/\gamma^* +\geq 1$ jet cross-section ratio. Here,
CTEQ5L~\cite{cteq5} PDFs are used, and the
factorization and renormalization scales are set to
$\mu_{F}=\mu_{R}=M_{Z}$.  The {\mcfm} and ME-PS predictions
are generally in good agreement with the data. {\pythia}
predicts fewer events at high jet multiplicity because of missing
higher order contributions at the hard-scatter level.

\begin{figure}[!]
\includegraphics[scale=0.46]{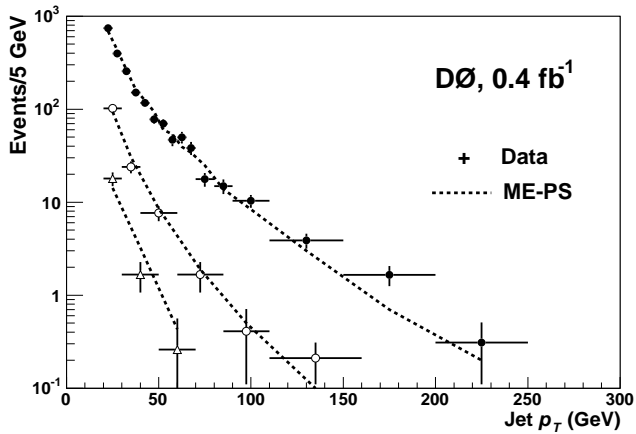}
\caption{\label{fig:fig_jets} Comparison between data and theory (ME-PS)
for the highest $p_T$ jet distribution in the
$Z/\gamma^* + \geq 1$ jet sample (dark circles), for the second
highest $p_T$ jet distribution in the $Z/\gamma^* + \geq 2$ jet
sample (open circles), and for the third highest $p_T$ jet
distribution in the $Z/\gamma^* + \geq 3$ jet sample (open
triangles). The uncertainties on the data are only statistical. The MC
distributions are normalized to the data.}
\end{figure}

Figure \ref{fig:fig_jets} compares jet $p_T$ spectra of the
$n^{th}$ jet, $n = 1,2,3$, in $Z/\gamma^* + \geq n$ jet events to
the ME-PS MC predictions.  The MC events have been passed
through the full detector simulation, and the jet $p_T$ spectra
normalized separately to the data distributions. Good agreement can
be seen over a wide range of jet transverse momenta.

In summary, we have presented the first measurements of fully corrected ratios
of the $Z/\gamma^* +\geq n$ jet ($n=1-4$) production cross
sections to the total inclusive $Z/\gamma^*$ cross section in
$p\overline{p}$ collisions at $\sqrt{s}=1.96$~TeV. The measured
ratios were found to be in good agreement with
{\mcfm} and an enhanced leading-order matrix element prediction
with {\pythia}-simulated parton showering and hadronization.
{\pythia} simulations alone appear to exhibit a deficit in high jet
multiplicity events.

We thank S. Mrenna for providing us the ME-PS MC sample.
%
We thank the staffs at Fermilab and collaborating institutions, 
and acknowledge support from the 
DOE and NSF (USA);
CEA and CNRS/IN2P3 (France);
FASI, Rosatom and RFBR (Russia);
CAPES, CNPq, FAPERJ, FAPESP and FUNDUNESP (Brazil);
DAE and DST (India);
Colciencias (Colombia);
CONACyT (Mexico);
KRF and KOSEF (Korea);
CONICET and UBACyT (Argentina);
FOM (The Netherlands);
PPARC (United Kingdom);
MSMT (Czech Republic);
CRC Program, CFI, NSERC and WestGrid Project (Canada);
BMBF and DFG (Germany);
SFI (Ireland);
The Swedish Research Council (Sweden);
Research Corporation;
Alexander von Humboldt Foundation;
and the Marie Curie Program.
%


\end{document}